\documentclass[aps,pra,twocolumn,showpacs,groupedaddress,amsmath,amssymb]{revtex4}

\usepackage{graphicx}
\graphicspath{{b:/officestuff/papers/nonlin_LRC/figures/}}

\usepackage{color}
\usepackage{comment}
\usepackage{bm}
\usepackage{latexsym}

\begin{document}

\title{Experimental realization of Floquet ${\cal PT}$-symmetric systems}
\author{Mahboobeh Chitsazi, Huanan Li, F. M. Ellis, Tsampikos Kottos} 
\affiliation{Department of Physics, Wesleyan University, Middletown, CT-06459, USA}
\date{\today}

\begin{abstract}
We provide an experimental framework where periodically driven ${\cal PT}$-symmetric systems can be investigated. 
The set-up, consisting of two UHF oscillators coupled by a time-dependent capacitance, demonstrates a cascade of 
${\cal PT}$-symmetric broken domains bounded by exceptional point degeneracies. These domains are analyzed and 
understood using an equivalent Floquet frequency lattice with local ${\cal PT}$-symmetry. Management of these 
${\cal PT}$-phase transition domains is achieved through the amplitude and frequency of the drive.
\end{abstract}

\pacs{05.45.-a, 42.25.Bs, 11.30.Er}
\maketitle


{\it Introduction -} Non-Hermitian Hamiltonians ${\cal H}$ which commute with the joint parity-time (${\cal PT}$) symmetry 
might have real spectrum when some parameter $\gamma$, that controls the degree of non-hermiticity, is below a critical 
value $\gamma_{\cal PT}$ \cite{BB98}. In this parameter domain, termed {\it exact} ${\cal PT}$-phase the eigenfunctions of 
${\cal H}$ are also eigenfunctions of the ${\cal PT}$-symmetric operator. In the opposite limit, coined the {\it broken} ${\cal 
PT}$-phase, the spectrum consists (partially or completely) of pairs of complex conjugate eigenvalues while the eigenfunctions 
cease to be eigenfunctions of the ${\cal PT}$ operator. The transition point $\gamma=\gamma_{\cal PT}$ shows all the characteristic 
features of an exceptional point (EP) singularity where both eigenfunctions and eigenvalues coalesce. 

Although originally the interest on ${\cal PT}$-symmetric systems was driven by a mathematical curiosity \cite{BB98}, during 
the last five years the field has blossomed and many applications in areas of physics, ranging from optics \cite{MGCM08,L09a,
RMGCSK10,ZCFK10,S10,SXK10,L10b,FXFLOACS13,GSDMVASC09,RCKVK12,CJHYWJLWX14,POLMGLFNBY14,HMHCK14,LT13,
RLKKKV12,LZOPJLLY16,PORYLMBNY14}, matter waves \cite{CW12,GKN08} and magnonics \cite{LKS14,GV16} to acoustics 
\cite{ZRSZZ14,FSA14,SDCCRWZ16} and electronics \cite{SLZEK11,CKSK13,BFBRCEK13}, have been proposed and experimentally 
demonstrated \cite{RMGCSK10,FXFLOACS13,GSDMVASC09,CJHYWJLWX14,POLMGLFNBY14,HMHCK14,LZOPJLLY16,PORYLMBNY14,
FSA14,SDCCRWZ16,SLZEK11}. Importantly, the existence of the ${\cal PT}$ phase transition and specifically of the EP singularity 
played a prominent role in many of these studies, and subsequent technological applications.

Though the exploitation of ${\cal PT}$-symmetric systems has been prolific, most of the attention has been devoted to static 
(i.e. time-independent) potentials. Recently, however, a parallel activity associated with time-dependent ${\cal PT}$-symmetric 
systems has started to attract increasing attention \cite{WKP10,M11,GMC12,VL13,LHZQXKL13,TL14,KZ14,AMK14,LZXZWL14,
JMDP14,WZHZZ16}. The excitement for this line of research stems from two reasons: the first one is fundamental and it is 
associated with the expectation that new pathways in the ${\cal PT}$-arena can lead to new exciting phenomena. This expectation 
is further supported by the fact that the investigation of time-dependent Hermitian counterparts led to a plethora of novel 
phenomena-- examples include Rabi oscillations \cite{SZ97}, Autler-Townes splitting \cite{AT55}, dynamical localization 
\cite{DK86}, dynamical Anderson localization \cite{CIS81}, and coherent destruction of tunneling \cite{GDJH91,VOCFLL07} 
(for a review see \cite{GH98}). The second reason is technological and it is associated with the possibility to use 
driving schemes as a flexible experimental knob to realize new forms of {\it reconfigurable} synthetic matter \cite{LRG11,WSHG13}. 
Specifically, in the case of ${\cal PT}$-symmetric systems one hopes that the use of periodic driving schemes can allow for a 
management of the spontaneous ${\cal PT}$-symmetry breaking for {\it arbitrary values of the gain and loss parameter}. The 
basic idea behind this is that periodic driving can lead to coherent destruction of tunneling leading to a renormalization of 
the coupling and a consequent tailoring of the position of the EPs. Unfortunately, while there is a number of theoretical studies 
\cite{M11,LHZQXKL13,TL14,JMDP14} advocating for this scenario, there is no experimental realization of a time-dependent 
${\cal PT}$-symmetric set-up. 

\begin{figure}
\includegraphics[width=1\linewidth, angle=0]{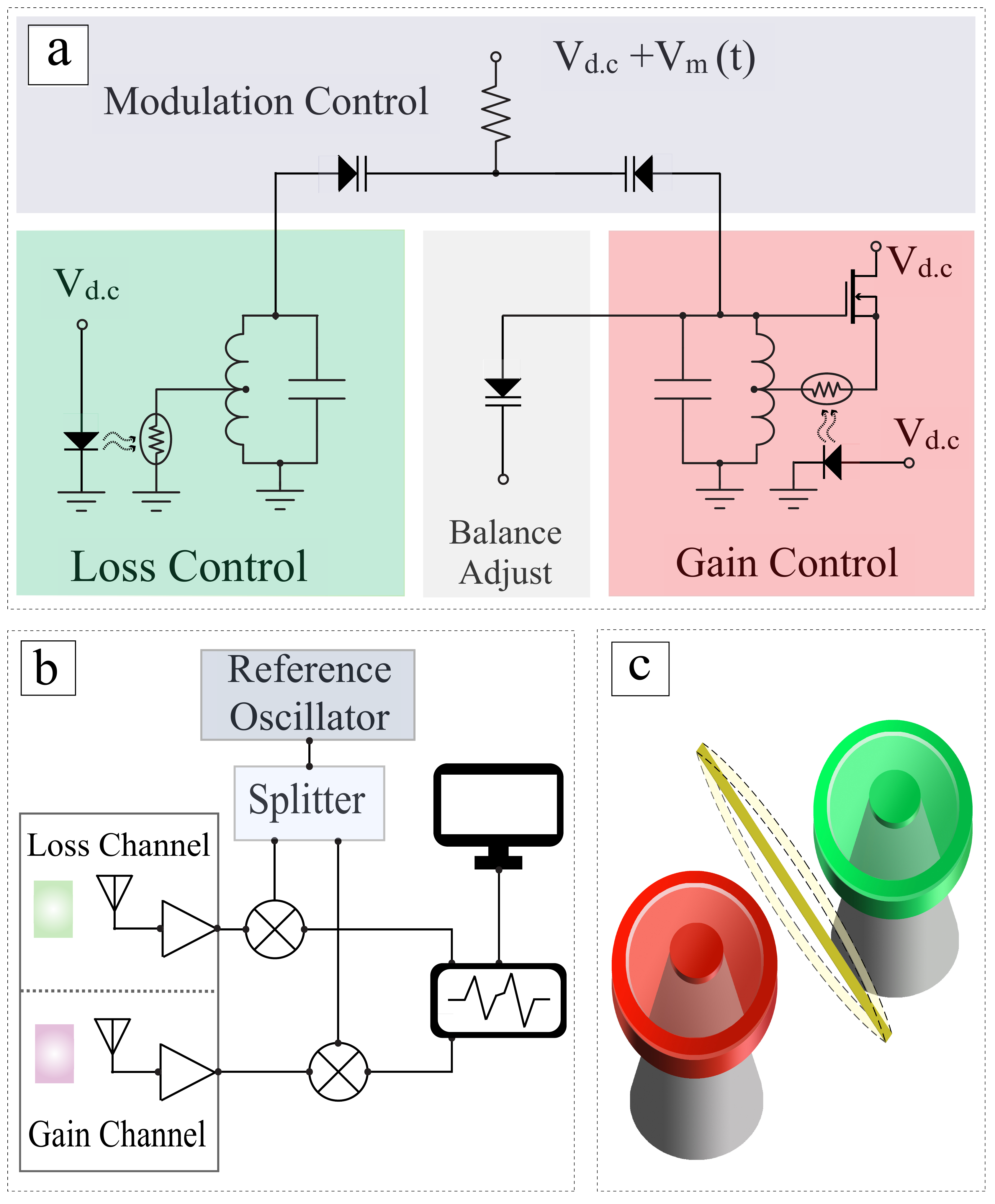}
\caption{(Color Online) (a) Experimental ${\cal PT}$ circuit with tuning and modulation control; (b) Signal control and analysis system; 
(c) An analogous mechanically modulated coupling (via a vibrating candilever) between two optical resonators, one with gain (red) and the
other with loss (green).   
}
\label{fig1}
\end{figure}

Here we provide such an experimental platform where periodically driven ${\cal PT}-$symmetric systems can be investigated. 
Our set-up (see Figs. \ref{fig1}a,b) consists of two coupled LC resonators with balanced gain and loss. The capacitance that 
couples the two resonators is parametrically driven with a  network of varactor diodes. We find that this driven ${\cal PT}$ 
system supports a sequence of spontaneous ${\cal PT}$-symmetry broken domains bounded by exceptional point degeneracies. 
The latter are analyzed and understood theoretically using an equivalent Floquet frequency lattice with local ${\cal PT}$-
symmetry. The position and size of these instability islands can be controlled through the amplitude and frequency of the driving 
and can be achieved, in principle, for arbitrary values of the gain/loss parameter. 

{\it Experimental set-up--} A natural frequency of $\omega_0/2\pi=235$ MHz was chosen as the highest frequency convenient for 
a simple implementation of electronic gain and loss. The $L=32$ nH inductors of Fig.~\ref{fig1}a are two-turns of 1.5 mm diameter 
Cu wire with their hot ends supported by their corresponding parallel $C=15$ pF on opposite sides of a grounded partition separating 
gain and loss compartments. Gain and loss (corresponding to effective parallel resistances $\mp R$) are directly implemented via 
Perkin-Elmer V90N3 photocells 
connecting the center turn of each inductor either directly to ground (loss side) or to a BF998 MOSFET following the $LC$ 
node (gain side). Thus as both photocells experience the same voltage drop, the loss side photocell extracts its current from 
the tap point while the gain side photocell injects its current into the tap point. The photocells are coupled to computer driven 
LEDs through 1 cm light-pipes for RF isolation. As the gain of the MOSFET is changed, its capacitance shifts slightly unbalancing
the resonators. A BB135 varactor is used to compensate for these changes.

The capacitance coupling network, implemented by similar varactors, is optimized for application of a modulation frequency 
in the vicinity of $4.6$ MHz modulation frequency while simultaneously providing the DC bias necessary for controlling the 
inter-resonator coupling $C_c$. 

Fig. \ref{fig1}b shows the remainder of the signal acquisition set-up. The excitation in each resonator is sensed by a small pickup 
loop attached to the input of a Minicircuits ZPL-1000 low noise amplifier. The gain and loss pick-up channels are then heterodyned 
to $\approx30$ MHz before being captured by a Tektronix DPO2014 oscilloscope. 

\begin{figure}
\includegraphics[width=1\linewidth, angle=0]{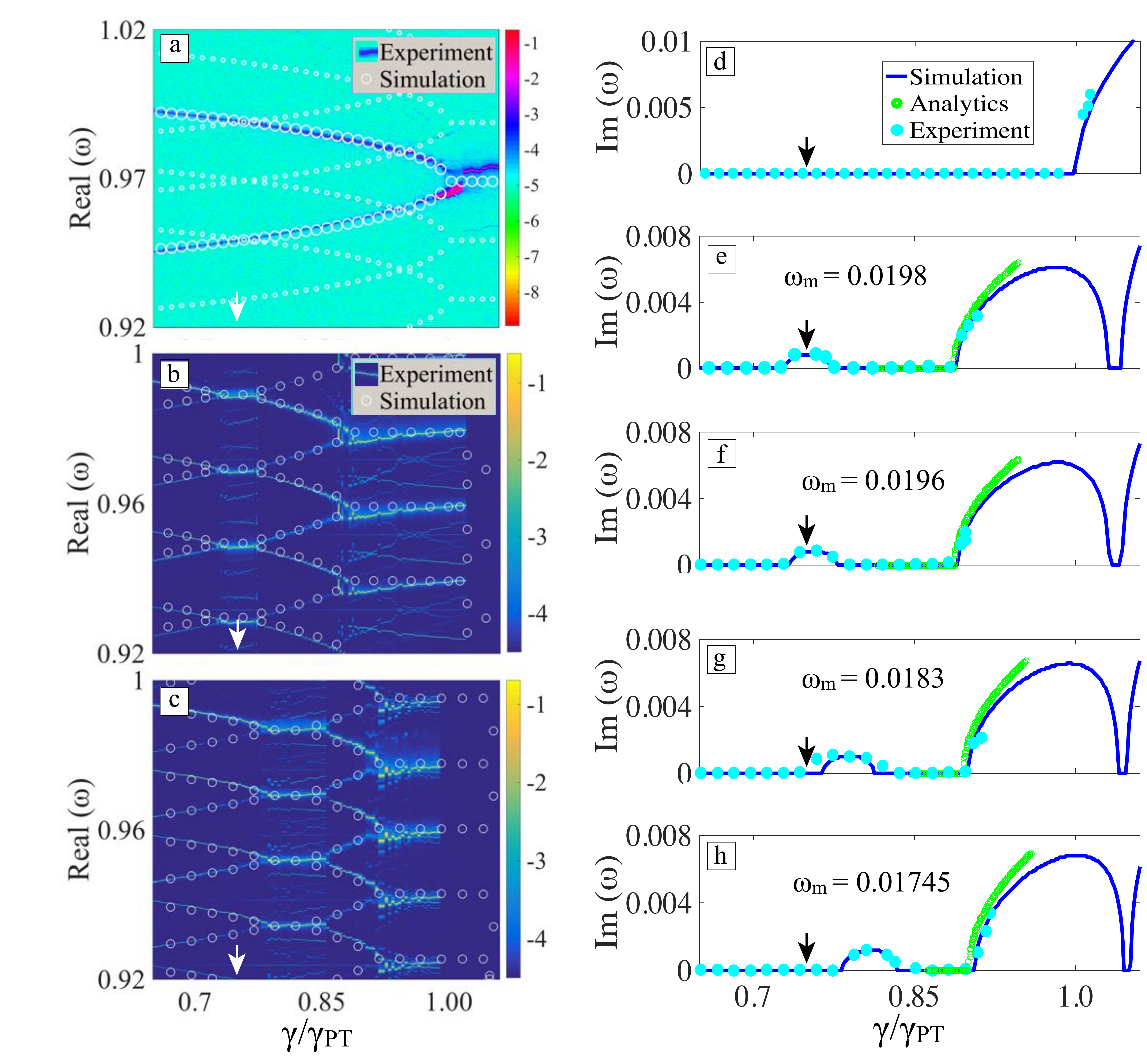}
\caption{(Color Online) Density plots for ${\cal R}e\left(\omega\right)$ of the RLC dimer of Fig. \ref{fig1}a,b : (a) Undriven dimer 
$\epsilon=0$. The white cycles represent the ladder $\omega_{1,2}+n\omega_m$ associated with the eigenfrequencies of $H_{F,0}$ for 
$\omega_m=0.0198$. The crossing points ``evolve'' to flat regions when the system is driven with (b) $\omega_m=0.0198$. (c) For different 
driving frequency $\omega_m=0.01745$ the flat regions shift to different $\gamma$-domains. The white circles in (b,c) represent the results
of the simulations \cite{Fnote}. Experimental measurements (aqua circles), numerics (blue line) and perturbation theory Eq. (\ref{qapprox})  (green 
circles) of ${\cal I}m\left(\omega\right)$  for (d) an undriven dimer and driven dimers with (e) $\omega_m=0.0198$ and (f) $\omega_m
=0.0196$ (g) $\omega_m=0.0183$ and (h) $\omega_m=0.01745$. In all cases the DC capacitance is $c_0=0.0671$ while the amplitude of 
the driving capacitance is $\epsilon=0.01$.
}
\label{fig2}
\end{figure}

\begin{figure}
\includegraphics[width=1\linewidth, angle=0]{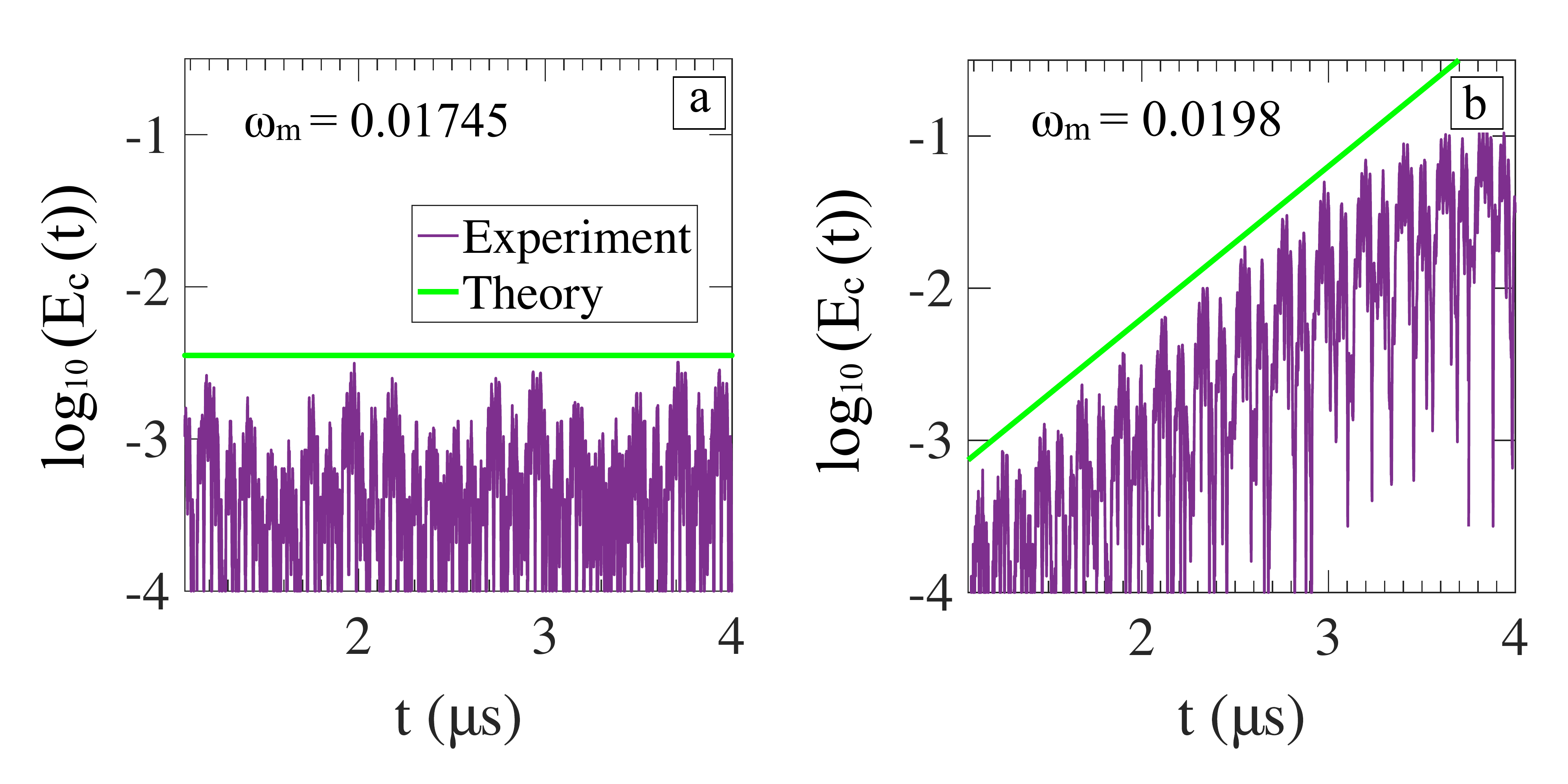}
\caption{(Color Online) Temporal evolution of normalized total capacitance energy $E_c(t)=\left(V_1^2(t)+V_2^2(t)\right)$ (in units of
Volt$^2$). The green lines indicate the theoretical predictions (from simulations) for the slope of the envelope for driving frequencies 
(a) $\omega_m= 0.01745$ and (b) $\omega_m=0.0198$. In both cases the value of $\gamma$ is the same (see black arrows in Fig. \ref{fig2}e,h). 
}
\label{fig3}
\end{figure}

The experimental unmodulated $\cal{PT}$ diagram, shown with the color-map in Fig. \ref{fig2}a, is matched to the theoretical results 
in order to calibrate both the resonator frequency balance and the gain/loss balance. The coupling is then modulated, directly comparing 
each calibrated point with and without the modulation. Signal transients are measured by pulsing the MOSFET drain voltage at approximately 
$1 kHz$ and capturing the resonator responses on both the gain and loss sides. The captured signals can be frequency-analyzed to obtain 
the modulated (or unmodulated) spectrum, see Fig. \ref{fig2}. Time transients can also be directly analyzed to reveal the time-domain dynamics,
see Fig. \ref{fig3}. Close attention has to be paid to avoid saturation of any of the components in the signal pick-up chain.
 
{\it Theoretical considerations--} Using Kirchoff\textquoteright s laws, the dynamics for the voltages $V_{1}\left(V_{2}\right)$ 
of the gain (loss) side of the periodically driven dimer is:
\begin{equation}
\frac{d^{2}}{d\tau^{2}}V+A\frac{d}{d\tau}V+BV=0; \quad V\equiv(V_{1},V_{2})^T,
\label{main}
\end{equation}
where $\tau=\omega_{0}t$ is the rescaled time, $\omega_{0}=\frac{1}{\sqrt{LC}}$ and 
\begin{eqnarray}
\label{ABmatrix}
A&=&\frac{1}{1+2c}\begin{bmatrix}-\gamma\left(1+c\right)
+2\stackrel{\centerdot}{c} & \gamma c-2\stackrel{\centerdot}{c}\\
-\gamma c-2\stackrel{\centerdot}{c} & \gamma\left(1+c\right)+2\stackrel{\centerdot}{c}
\end{bmatrix}\nonumber\\
B&=&\frac{1}{1+2c}\begin{bmatrix}1+c+\stackrel{\centerdot\centerdot}{c} & c-\stackrel{\centerdot\centerdot}{c}\\
c-\stackrel{\centerdot\centerdot}{c} & 1+c+\stackrel{\centerdot\centerdot}{c}
\end{bmatrix} 
\end{eqnarray}
Above $\gamma=R^{-1}\sqrt{L/C}$ is the rescaled gain/loss parameter, and $\stackrel{\centerdot}{c}$ ($\stackrel{\centerdot
\centerdot}{c}$) denotes the first (second) derivative of the scaled capacitive coupling $c\equiv\frac{C_{c}}{C}=c_{0}+
\varepsilon\cos\left(\omega_m\tau\right)$ with respect to the scaled time $\tau$. Equation (\ref{main}) is invariant under joint 
parity ${\cal P}$ and time ${\cal T}$ operations, where ${\cal T}$ performs the operation $\tau\rightarrow -\tau$ and ${\cal P}$ 
is the Pauli matrix $\sigma_x$.

The eigenfrequencies $\omega_{\alpha}\:(\alpha=1,2)$ of system Eq. (\ref{main}) in the absence of driving are given as 
\begin{align}
\label{undrivenfreq}
\omega_{\alpha}= & \frac{1}{2\sqrt{1+2c_{0}}}\left(\sqrt{\gamma_{c}^{2}-\gamma^{2}}+\left(-1\right)^{\alpha}\sqrt{\gamma_{PT}^{2}-\gamma^{2}}\right)
\end{align}
where the spontaneous ${\cal PT}$- symmetry breaking point and the upper critical point can be identified as $\gamma_{{\cal PT}}=\sqrt{1+2c_{0}}-1$ 
and $\gamma_{c}=\sqrt{1+2c_{0}}+1$ respectively, and they are both determined by the strength of the (capacitance) coupling 
between the two elements of the dimer. A parametric evolution of these modes, versus the gain/loss parameter $\gamma$, is 
shown in Fig. \ref{fig2}a where the open circles are Eq. (\ref{undrivenfreq}) and the color map shows the experimental results. 
We find that the spectrum is divided in two compact domains of exact ($\gamma<\gamma_{\cal PT}$) and broken ($\gamma>
\gamma_{\cal PT}$) ${\cal PT}$-symmetric domains.

In order to investigate the effects of driving we now turn to the Floquet picture. A simple way to do this is by employing a 
Liouvillian formulation of Eq.~(\ref{main}). The latter takes the form 
\begin{align}
\label{schro}
\frac{d\psi}{d\tau}= & \mathcal{L}\psi,\quad\:\mathcal{L}=\begin{bmatrix}0 & I_{2}\\
-B & -A
\end{bmatrix}, \,\,\,\,\psi=\left(\begin{matrix}V\\\stackrel{\centerdot}{V}\end{matrix}\right)
\end{align}
and allows us to make equivalences with the time-dependent Schr\"odinger equation by identifying a non-Hermitian effective Hamiltonian 
$H_{eff}=\imath {\mathcal L}$.  

The general form of the solution of Eq.~(\ref{schro}) is given by Floquet's theorem which in matrix notation reads  $F\left(\tau\right)=
\Phi\left(\tau\right)e^{-\imath Q\tau}$ with $\Phi\left(\tau+\frac{2\pi}{\omega_m}\right)=\Phi\left(\tau\right)$,  
$Q$ a Jordan matrix and $F(\tau)$ a $4\times 4$ matrix consisting of four independent solutions of Eq. (\ref{schro}).
The eigenvalues of $Q$ are 
the characteristic exponents (quasi-energies) which determine the stability properties of the system: namely the system 
is stable (exact ${\cal PT}$ phase) if all the quasi-energies are real and it is unstable (broken ${\cal PT}$ phase) otherwise.
We can evaluate the quasi-energies by constructing the evolution operator $U\left(\tau,0
\right)=F(\tau)F^{-1}(0)$ via numerical integration of Eq.~(\ref{schro_eq}) (or of Eq. (\ref{main})). Then the quasi-energies are 
the eigenvalues of $\frac{1}{-\imath2\pi/ \omega_m}\ln U\left(\tau=2\pi/\omega_m,0\right)$. 

In Figs.~\ref{fig2}a-h we report our numerical findings together with the experimentally measured values of the quasi-energies 
versus the gain/loss parameters. Figs.~\ref{fig2}a,d show the unmodulated situation. Fig.~\ref{fig2}b,e show the behavior at 
modulation frequency $\omega_m=0.0198$ and modulation amplitude $\epsilon=0.01$. Finally, Figs~\ref{fig2}c,e-h show the 
evolution of the spectrum with a small change in modulation frequency $\omega_m$ for fixed $\epsilon$. Only one of the real 
part color-maps, $\omega_m=0.01745$, is shown.

We find several new features in the spectrum of the driven ${\cal PT}$-symmetric systems. The first difference with respect to 
the undriven case is the existence of a cascade of domains for which the system is in the broken ${\cal PT}$-phase. These 
domains are identified by the flat regions, seen in Figs. \ref{fig2}b,c where the real parts of eigenfrequencies have merged, or by 
the emerging non-zero imaginary parts shown in Figs. \ref{fig2}e-h. The size and position of these unstable ``bubbles" are directly 
controlled by the values of the driving amplitude $\epsilon$, compare Figs.~\ref{fig2}a,d with Figs. \ref{fig2}b,e or by the driving 
frequency $\omega_m$, compare Figs. \ref{fig2}b,e with Figs. \ref{fig2}c,h. The bubbles are separated by $\gamma$-domains where 
the system is in the exact (stable) ${\cal PT}$-phase. The transition between stable and unstable domains occurs via a typical EP 
degeneracy (notice the square -root singularities in Figs. \ref{fig2}d-h).  

A theoretical understanding of the spectral metamorphosis from a single compact exact/broken 
phase to multiple domains of broken and preserved ${\cal PT}$-symmetry, as $\epsilon$ increases from zero, is achieved 
by utilizing the notion of Floquet Hamiltonian $H_{F}$. To this end, we first introduce a time-dependent similarity transformation 
${\cal R}$, which brings the effective Hamiltonian matrix that dictates the evolution in the Schr\"odinger-like equation to a 
symmetric form. We shall see below that the symmetric form is inherited 
in the Floquet Hamiltonian (up to the first order perturbation in $\epsilon$ and  $\omega_m\sim O(\varepsilon)$). Consequently the bi-orthogonal Floquet eigenmodes 
of the unperturbed Floquet matrix are transpose of each other -- a property that greatly simplifies the analytical process for 
the evaluation via first-order theory of the Floquet eigenmodes. This transformation takes the form \cite{SLZEK11}:
\begin{align}
{\cal R}\left(\tau\right)= & \begin{bmatrix}1 & 1 & \imath\sqrt{\beta} & -\imath\sqrt{\beta}\\
\sqrt{1+2\stackrel{\centerdot\centerdot}{c}} & -\sqrt{1+2\stackrel{\centerdot\centerdot}{c}} & \imath & \imath\\
-\sqrt{1+2\stackrel{\centerdot\centerdot}{c}} & \sqrt{1+2\stackrel{\centerdot\centerdot}{c}} & \imath & \imath\\
1 & 1 & -\imath\sqrt{\beta} & \imath\sqrt{\beta}
\end{bmatrix}
\end{align}
and allows us to bring Eq.~(\ref{schro}) to the following Schr\"odinger-like form 
\begin{align}
\imath\frac{d}{d\tau}\tilde{\psi}= & {\tilde H}\tilde{\psi};\quad {\tilde H}\equiv {\cal R}H_{eff}{\cal R}^{-1}-\imath {\cal R}
\frac{d}{d\tau}{\cal R}^{-1}
\label{schro_eq}
\end{align}
which dictates the evolution of the transformed state $\tilde{\psi}={\cal R}\psi$. The matrix ${\tilde H}$ satisfies the relation 
${\tilde H}={\tilde H}^T$, i.e. it is transposition symmetric, and has the form  
\begin{align}
{\tilde H}= & \begin{bmatrix}-\frac{3}{2}\frac{\imath\stackrel{\centerdot}{c}}{\beta} & c_{+}+\frac{\imath\gamma}{2\sqrt{\beta}} & c_{-}+\frac{\imath\gamma}{2\sqrt{\beta}} & \frac{3}{2}\frac{\imath\stackrel{\centerdot}{c}}{\beta}\\
c_{+}+\frac{\imath\gamma}{2\sqrt{\beta}} & \frac{\imath c^{(3)}}{2(1+2\stackrel{\centerdot\centerdot}{c})} & -\frac{\imath c^{(3)}}{2(1+2\stackrel{\centerdot\centerdot}{c})} & c_{-}-\frac{\imath\gamma}{2\sqrt{\beta}}\\
c_{-}+\frac{\imath\gamma}{2\sqrt{\beta}} & -\frac{\imath c^{(3)}}{2(1+2\stackrel{\centerdot\centerdot}{c})} & \frac{\imath c^{(3)}}{2(1+2\stackrel{\centerdot\centerdot}{c})} & c_{+}-\frac{\imath\gamma}{2\sqrt{\beta}}\\
\frac{3}{2}\frac{\imath\stackrel{\centerdot}{c}}{\beta} & c_{-}-\frac{\imath\gamma}{2\sqrt{\beta}} & c_{+}-\frac{\imath\gamma}
{2\sqrt{\beta}} & -\frac{3}{2}\frac{\imath\stackrel{\centerdot}{c}}{\beta}
\end{bmatrix},
\label{Hmatrix}
\end{align}
where $\beta=1+2c, c_{\pm}=\frac{1}{2}\pm\frac{1}{2}\frac{\sqrt{1+2\stackrel{\centerdot\centerdot}{c}}}{\sqrt{\beta}}$ and 
$c^{(3)}$ denotes 
the third derivative of the capacitive coupling with respect to the scaled time $\tau$. We can easily show that $\tilde{\cal P}{\cal T}
{\tilde H}\tilde{\cal P}{\cal T}={\tilde H}$ where ${\tilde {\cal P}}=\begin{bmatrix}0 & \sigma_{x}\\\sigma_{x} & 0\end{bmatrix}$ 
and the time-reversal operator is defined as ${\cal T}: \tau\rightarrow -\tau, \imath \rightarrow -\imath$.

We are now ready to utilize the notion of Floquet Hamiltonian $H_{F}$ whose components are  given by
\begin{align}
\left\langle \alpha,n\right|H_{F}\left|\beta,l\right\rangle = & {\tilde H}_{\alpha\beta}^{\left(n-l\right)}+n\omega_m\delta_{\alpha\beta}\delta_{nl},
\label{H_F}
\end{align}
where the subscripts $\alpha,\beta=1,2,3,4$ label the components of ${\tilde H}$, see Eq. (\ref{Hmatrix}), $n,l$ are any 
integers and ${\tilde H}_{\alpha\beta}^{\left(n\right)}=\frac{1}{2\pi/\omega_m}\int_{0}^{2\pi/\omega_m}{\tilde H}_{\alpha\beta}
\left(\tau\right)e^{-\imath n\omega_m\tau}d\tau$. In this picture the quasi-energies are the eigenvalues of 
the Floquet Hamiltonian $H_{F}$. Note, that Eq. (\ref{H_F}) defines a lattice model with complex connectivity given by the 
off-diagonal elements of $H_F$ and an on-site gradient potential $n\omega_m$. These type of lattices (though in real space) 
have been recently investigated in the frame of waveguide photonics and have been shown to generate Bloch-Zener type of 
dynamics \cite{BLEK15}.  Thus our set-up can be used as a user-friendly experimental framework for the investigation of 
dynamics in such complicated lattices where the lattice complexity enters via a designed driving scheme.

Within the first order approximation to the strength of the driving amplitude $\varepsilon$ and the modulation frequency  $\omega_m\sim O(\varepsilon)$,
 the Floquet Hamiltonian is symmetric 
and takes the block-tridiagonal form $H_{F}=H_{F,0}+\varepsilon H_{F,1}+O\left(\varepsilon^{2}\right)$ where $\left\langle n
\right|H_{F,0}\left|n\right\rangle ={\tilde H}^{\left(0\right)}\left|_{\varepsilon=0}\right.+n\omega_m I_{4}$ consists of the diagonal 
blocks of $H_{F}$ while $\left\langle n+1\right|H_{F,1}\left|n\right\rangle =\left\langle n\right|H_{F,1}\left|n+1\right\rangle 
=X$ consist of off-diagonal blocks of $H_{F}$. The $4\times 4$ matrix $X$ has the form \cite{note1}
\begin{align}
X=\frac{\imath}{4\left(1+2c_{0}\right)^{3/2}} & \begin{bmatrix}0 & \imath-\gamma & -\imath-\gamma & 0\\
\imath-\gamma & 0 & 0 & -\imath+\gamma\\
-\imath-\gamma & 0 & 0 & \imath+\gamma\\
0 & -\imath+\gamma & \imath+\gamma & 0
\end{bmatrix}.
\label{Xmatrix}
\end{align}

Next we proceed with the analytical evaluation of the quasi-energies. First, we neglect the off-diagonal block matrices $H_{F,1}$ 
and diagonalize $H_{F,0}$. To this end, we construct a similarity transformation $P_{0}^{-1}H^{\left(0\right)}\left|_{\varepsilon=0}
\right.P_{0}=diag\left\{ \omega_{2},\omega_{1},-\omega_{1},-\omega_{2}\right\} $. Correspondingly the eigenvalues of $H_{F,0}$ 
are simply $\left\{ \omega_{2}+n\omega_m,\omega_{1}+n\omega_m,-\omega_{1}+n\omega_m,-\omega_{2}+n\omega_m\right\}$ 
where $n$ is an integer i.e. the spectrum resembles a ladder of step $\omega_m$ and basic unit associated with the eigenfrequencies 
of the undriven dimer Eq. (\ref{undrivenfreq}). The resulting ladder spectrum (white circles) is shown in Fig. \ref{fig2}a versus the 
gain/loss parameter $\gamma$. Level crossing occurs at some specific values of $\gamma^{\left(j\right)}$, i.e., $\omega_{2}
\left|_{\gamma^{\left(j\right)}}\right.=\omega_{1} \left|_{\gamma^{\left(j\right)}}\right.+j\omega_m$. When the driving 
amplitude $\varepsilon$ is turned on, the crossing points evolve to broken ${\cal PT}$-symmetry domains with respect 
to gain/loss parameter $\gamma$ (see Figs. \ref{fig2}b,c). Specifically the real part of the eigenfrequencies become degenerate 
for a range of $\gamma$-values around $\gamma^{(j)}$, Fig. \ref{fig2}b, while an instability bubble emerges for the imaginary 
part -- see Fig. \ref{fig2}e where numerical (blue solid lines) and experimental data (filled aqua circles) agree nicely with one another. 
The transition points from stable to unstable domains have all the characteristic features of an EP i.e. a square root singularity of 
the spectrum and degeneracy of the eigenmodes (not shown). 

To understand this phenomenon, we now consider the effect of the additional off-diagonal term $\varepsilon H_{F,1}$. For simplicity, 
we focus on the unstable region around the crossing point at $\gamma^{\left(1\right)}$. Application of degenerate perturbation 
theory to the nearly degenerate levels $\omega_{2}$ and $\omega_{1}+\omega_m$ gives 
\begin{align}
\omega= & \frac{\left(\omega_{2}+\omega_{1}+\omega_m\right)\pm\sqrt{\left(\omega_{2}-\omega_{1}-\omega_m\right)^{2}
+4\varepsilon^{2}\tilde{X}_{12}\tilde{X}_{21}}}{2},
\end{align}
where $\tilde{X}=P_{0}^{-1}XP_{0}$ and the subscripts indicate the corresponding matrix components. Explicitly, around the EP, $\omega$ 
can be written as
\begin{align}
\label{qapprox}
\mathrm{Re}\left(\omega\right)\approx\omega_{2}\left|_{\gamma^{\left(1\right)}}\right.; \,\, 
\mathrm{Im}\left(\omega\right)= & \pm C_{m}\sqrt{\gamma-\gamma_0},\,\, \gamma>\gamma_{0}
\end{align}
which has the characteristic square-root singularity of EP degeneracies. The constant $C_m$ takes the form 
\begin{widetext}
\begin{align}
C_{m}= & \frac{1}{2}\sqrt{\frac{1}{\gamma_{PT}^{2}-\gamma_{0}^{2}}\frac{\varepsilon\gamma_{0}}{1+2c_{0}}
\left[\left(\frac{2}{\sqrt{1+2c_{0}}}-\frac{\omega_m}{\sqrt{\gamma_{PT}^{2}-\gamma_{0}^{2}}}\right)\gamma_{0}+
\frac{\varepsilon}{2\left(1+2c_{0}\right)}\right]}
\end{align}
\end{widetext}
and $\gamma_{0}$ is the solution of the equation $\left(\omega_{2}-\omega_{1}-\omega_m\right)^{2}+4\varepsilon^{2}
\tilde{X}_{12}(\gamma)\tilde{X}_{21}(\gamma)=0$
(see Eq. (\ref{Xmatrix})) \cite{note2}. From Eq. (\ref{qapprox}) we clearly see that both $\omega_m,\epsilon$ are responsible for a 
renormalization of the coupling between the two levels (compare with Eq. (\ref{undrivenfreq})).

Predictions (\ref{qapprox}) are in agreement with the numerical and experimental data (see green line in Fig. \ref{fig2}e-h). Higher
orders of EPs $\gamma^{(j)}$ can be analyzed in a similar manner. In this case, however, the small size of the instability bubbles 
require to take into consideration higher order perturbation theory corrections in order to describe the quasi-energies $\omega$. We 
have, nevertheless tested the applicability of this scheme via numerical evaluation of the higher order corrections.

The management of the exact and broken ${\cal PT}$ symmetry phase, either via the driving amplitude $\epsilon$ or via the 
frequency $\omega_m$, has also direct implications to the dynamics of the system. In Fig.\ref{fig3} we report the total capacitance 
energy of the dimer for the same $\gamma=0.0483$ and $\epsilon=0.01$ values but different driving frequencies $\omega_m=
0.01745$ (left) and $\omega_m=0.0198$ (right). In the latter case the energy grows exponentially with a rate given
by the imaginary part of the eigenfrequencies (see Fig. \ref{fig2}e) while in the former we have an oscillatory (stable) dynamics (see Fig. 
\ref{fig2}h).

{\it Conclusions--} We have experimentally demonstrated that ${\cal PT}$-symmetric systems containing periodically driven 
components are capable of controlling the presence, strengths, and positions of multiple exact-phase domains, bounded by corresponding exceptional 
points. The generic behavior is well described by a perturbative analysis of the Floquet Hamiltonian, and opens up new directions 
of exceptional point management in a variety of electronic, mechanical or optimechanical applications, similar to the example shown 
in Fig. \ref{fig1}c.


{\it Acknowledgments --} 
This research was partially supported by an AFOSR grant No. FA 9550-10-1-0433, and by NSF grants ECCS-1128571 and DMR-1306984.


\begin{thebibliography}{99}
\bibitem{note1} It turns out that for our specific experimental parameters $\omega_m\sim O(\varepsilon)$. Nevertheless in our calculation 
scheme we include the terms $n\omega_m I_{4}$ in the $H_{F,0}$ part since they are analytically easy to deal with.
\bibitem{note2} In the specific case of $\gamma^{(1)}$ and after taking into consideration our experimental case  $\gamma_{0}\rightarrow
\gamma_{\cal PT}$ and $\gamma_{\cal PT}\rightarrow0$ we can explicitly solve and have that $
\gamma_{0}\approx \gamma_{\cal PT}\left(1-\left(\frac{\sqrt{2}\omega_m+\sqrt{2\omega_m^{2}+\varepsilon\left(4\gamma_{PT}+\varepsilon\right)}}{4\gamma_{PT}+\varepsilon}\right)^{2}\right)
$.
\bibitem{Fnote}Note that the real parts in Figs. \ref{fig2}b,c include a mode translated from the negative frequency domain
of the Floquet latice that is not shown due to its practical insignificance and enhanced sensitivity to drive frequency.
\end{thebibliography}
\end{document}